\documentclass{article}

\usepackage[T1]{fontenc}
\usepackage[utf8]{inputenc}
\usepackage{authblk}
\usepackage{amsmath,amsthm,amsfonts}
\usepackage[inline]{enumitem}
\usepackage{pgfplots}
\usepackage{tikz}
\usetikzlibrary{calc}

\newcommand{\grad}[0]{\nabla}
\renewcommand{\div}[0]{\nabla \cdot}
\newcommand{\lap}[0]{\nabla^2}
\newcommand{\ddt}[1]{\frac{\partial #1}{\partial t}}
\newcommand{\vel}[0]{\vec{u}}
\newcommand{\filter}[1]{\langle #1 \rangle}
\newcommand{\rav}[1]{\overline{#1}}
\newcommand{\ros}[1]{#1^\prime}
\newcommand{\fav}[1]{\tilde{#1}}

\newcommand{\dd}[2][x]{\frac{\partial#2}{\partial #1}}

\title{Diluted-dispersed mass transfer within an AWE}

\author{N.Valle}

\affil{
    Process \& Energy Department,
    Delft University of Technology,
    Leegwaterstraat 39,
    2628 CB Delft,
    The Netherlands
}

\date{}

\begin{document}

\maketitle

\begin{abstract}
    The goal of this document is describe the multiphase transfer processes 
    describing the bubble dynamics of a water electrolyzer.
    The motivation is to describe the dilute-dispersed mass transfer within and 
    Alkaline Water Electrolyzer.
    Special emphasis is put on the mathematical formulation.
    The presentation starts by posing the governing equations
    and their dimensionless counterpart.
    By filtering the equations, the two-fluid model is presented
    along with the need to sub-scale and wall models.
    To the later aim,
    boundary layer equations are introduced.
    By reviewing self-similiarity transformations,
    the analysis of Blasius, Ostrach and Sparrow
    is reviewed for Prandtl's boundary layer equations;
    along with that of Leveque.
\end{abstract}

\section{Governing equations}
\textbf{Conservation laws}

We begin with the formulation of mass, momentum and species conservation:

\begin{align}
    \label{eqn:conservation-mass}
    \ddt{\rho} + \div \rho \vel                     &= 0 \\
    \label{eqn:conservation-momentum}
    \ddt{\rho \vel} + \div \rho \vel \otimes \vel   &=
            -\grad p + \div \tau + \rho\vec{g}\\
    \label{eqn:conservation-species}
    \ddt{c} + \div c \vel                           &= D\div f + k c
\end{align}

\textbf{Boundary conditions}

Boundaries consist of the channel walls along with the input and output regions 
of the channel.

\begin{align}
    \label{eqn:boundary-momentum-input}
    p\rvert_{i}                             &= p_0  &
    p\rvert_{i}                             &= 0    &
    \vel\rvert_{o}\hat{t}_o                 &= 0    &
    \vel \rvert_{w}                         &= 0    \\
    \label{eqn:boundary-species-input}
    c \rvert_{i}                            &= c_s  &
    D \grad c \rvert_{o} \cdot \hat{t}_{o}  &= 0    &
                                            &       &
    D \grad c \rvert_{w} \cdot \hat{n}_{w}  &= k c
\end{align}

\textbf{Local instant formulation}

Now, introducing the marker function $\phi$,
we proceed to the splitting of the continuum equations between continuum (i.e., 
liquid) and dispersed (i.e., gas) phases.

\begin{align}
    \label{eqn:conservation-mass-lic}
    \ddt{\phi_k \rho} + \div \phi_k \rho \vel                   &= m_k \\
    \label{eqn:conservation-momentum-lic}
    \ddt{\phi_k \rho \vel} + \div \phi_k \rho \vel \otimes \vel &=
            -\grad \phi p  + \mu \div \phi_k S + \phi_k c \vec{g} + M_k\\
    \label{eqn:conservation-species-lic}
    \ddt{\phi_k c} + \div \phi_k c \vel                         &=
            D\lap \phi_k c + k \phi_k c
\end{align}
where $m_k$ corresponds to the mass transfer between phases,
while $M_k$ corresponds to the momentum transfer.

\begin{align}
    \dot{m}_k   &=  \rho_k (\vel_\Gamma - \vel) \cdot \hat{n}_\Gamma 
                    \delta_\Gamma\\
    M_k         &=  \dot{m}_k \cdot \vel + \sigma \hat{n} \delta_\Gamma
\end{align}

\textbf{Incompressible}

\begin{equation}
    \vel = 0
    \label{eqn:incompressible}
\end{equation}

\textbf{Newtonian}

We will assume that the fluid is Newtonian, and so that the relationship between 
strain ($S$) and stress ($\tau$) is linear:
\begin{equation}
    \tau = \mu S
    \label{eqn:Newtonian}
\end{equation}

\textbf{Boussinesq}
While both density and viscosity are functions of the local supersaturation,
we will adopt here the Boussinesq hypothesis.
This is reasonable as far as the concentrations are close to saturation.
Consequently, we will consider them constant everywhere except for the 
formulation of the buoyancy term,
where we approximate it with a 2nd order Taylor expansion:
\begin{equation}
    \rho(c) \approx \rho_s + \frac{\partial \rho}{\partial c} (c - c_s)
    \label{eqn:Boussinesq}
\end{equation}
Nonetheless,
since the dispersed phase is composed of very small bubbles,
whose concentration gradients \textbf{within} the bubble itself are expected to 
be very small,
buoyancy effects \textbf{inside} the bubbles have little room to develop and so 
are neglected.
Actually, we consider concentration to be constant within the dispersed phase 
and equal to $c_s$.
Note, however,
that density differences \textbf{across} the bubble-liquid interface do induce  
strong buoyancy effects on the bubbles relative to the liquid phase,
which are effectively included.
Consequently,
we will only introduce the Boussinesq hypothesis for the liquid phase
(i.e., $k = l$).

\textbf{Final form}

By referencing the buoyancy terms to $\rho_{s}$, we finally obtain:
\begin{align}
    \label{eqn:governing-mass}
    \ddt{\phi_k}      + \div \phi_k \vel                        &=
            \dot{m}_k                      \\
    \label{eqn:governing-momentum-liq}
    \ddt{\phi_l \rho \vel} + \div \phi_l \rho \vel \otimes \vel &=
            - \grad \phi_l p + \div \phi_l \mu S
            + \phi_l \frac{\partial \rho}{\partial c} (c - c_s) \vec{g}
            + M_l                                                       \\
    \label{eqn:governing-momentum-gas}
    \ddt{\phi_g \rho \vel} + \div \phi_g \rho \vel \otimes \vel &=
            - \grad \phi_g p + \div \phi_g \mu S
            + \phi_g (\rho - \rho_l) \vec{g}
            + M_g                                                       \\
    \label{eqn:governing-species-liq}
    \ddt{\phi_l c} + \div \phi_l c u^i                          &=
            D\lap \phi_l c + k \phi_l c - \dot{s}_l \\
    \label{eqn:governing-species-liq}
    \phi_g c                                                    &= c_s
\end{align}
which consitutes the local instantious formulation of the governing equations.
We can see that liquid and gas phases exhibit a different nature of the buoyancy 
terms (concentration driven in the liquid, vs. density driven in the gas).
We also see that species transport is only relevant within the liquid phase.
Note the difference between the effect between species (and mass) transfer in 
both dispersed and continuos phases:
for the gas phase,
whose concentration is constant,
$\dot{s}_g$ contributes only to the expansion of the phase via the conservation 
of mass;
whereas for the liquid phase,
whose concentration is variable,
$\dot{s}_l$ both contributes to the shrinking of the gas phase and its dilution.

\subsection{Modeling of closure terms}

\textbf{Mass transfer coefficient}

While the physical mechanism of mass transfer is described by the diffusion of 
dissolved gas from the liquid into the bubble,
it is most of the time convenient to model such a flow by a mass transfer 
coefficient $k_b$

\begin{equation}
    D\nabla c \rvert_\Gamma = k_b (c-c_s) \rvert_\Gamma
    \label{eqn:mass-transfer-coefficient}
\end{equation}

such that we obtain the flux of species as:
\begin{equation}
    \dot{s}_k   =  \pm k_b (c-c_s) \delta_\Gamma
    \label{eqn:species-closure}
\end{equation}
which turns into a mass flux by using the molar density $N$
\begin{equation}
    \dot{m}_k   =  \pm N k_b (c-c_s) \delta_\Gamma
    \label{eqn:species-closure}
\end{equation}

\section{Dimensionless formulation}
We now introduce the general dimensionless variables for a buoyancy driven flow:
\begin{align*}
    x       &= L X                                              &
    t       &= t_{ref}          \tau= \frac{\rho L^2}{\mu} \tau &
    u       &= \frac{L}{t_{ref}}U   = \frac{\mu}{\rho L}    U   &
    p       &= \rho_0 u_{ref}^2 P   = \frac{\mu^2}{\rho L^2}P  \\
    c       &= c_s\left( \zeta + 1 \right)                      &
    \rho    &= \rho_0 \rho^*                                    &
    \mu     &= \mu_0 \mu^*                                      &
    D       &= D_0 D^*                                          \\
\end{align*}
where we have used phase $0$ as the reference values of the physical properties,
and introduced $\rho^*$, $\mu^*$ and $D^*$ as dimensionless properties.

With this formulation,
we can restate mass, species and momentum conservation in dimensionless form 
as:

\begin{align}
    \label{eqn:conservation-mass-lic-dimless}
    \ddt{\phi_k} + \div \phi_k \vel                     &=
            \pm \boxed{ \frac{N c_s}{\rho_k}}~
                \boxed{ \frac{k_b L \rho_0}{\mu}} \zeta \delta_\Gamma \\
    \label{eqn:conservation-momentum-lic-dimless-liq}
    \ddt{\phi U^i} + \div \phi U^i U^j                  &=
            -\grad \phi P + \div \phi S
            +\boxed{\frac{gL^3c_s\rho_s}{\mu_s}\frac{\partial \rho}{\partial c}}
             \zeta \vec{g}                                          \\
    \label{eqn:conservation-momentum-lic-dimless-gas}
    \ddt{\phi \rho^* U^i} + \div \phi \rho^* U^i U^j    &=
            -\grad \phi P + \div \phi \mu^* S
            +\boxed {\frac{\rho_l(\rho_g-\rho_l)gL^3}{\mu_0^2}}
             \rho^* \zeta \vec{g}                                   \\
    \label{eqn:conservation-species-lic-dimless}
    \ddt{\phi \zeta} + \div \zeta U^i                   &=
              \boxed{\frac{D\rho_s}{\mu_s}} D^* \lap \zeta
            + \boxed{\frac{k L^2 \rho_s}{\mu_s}} (\zeta + 1)
            + \boxed{\frac{k_b L \rho}{\mu}} \zeta
\end{align}

and also the boundary conditions
\begin{align}
    \label{eqn:boundaries-momentum-noslip-dimless}
    \frac{\mu}{\rho L} U \rvert_{wall} &= 0\\
    \label{eqn:boundaries-species-flux-dimless}
    \frac{D c_s}{L} \grad \zeta \rvert_{wall} \cdot \hat{n}_{wall} &= k c_s 
    (\zeta + 1)
\end{align}

Which can be rewritten most compactly by identifying the characteristic 
dimensionless numbers as follows:
\begin{align}
    \label{eqn:dimless-mass-lic-gas}
    \ddt{\phi_l} + \div \phi_l U                            &=
            - \frac{\rho_g}{\rho_l}
                \frac{Sh_b}{Sc} \zeta \delta_\Gamma         \\
    \label{eqn:dimless-mass-lic-liq}
    \ddt{\phi_g} + \div \phi_g U                            &=
            + \frac{Sh_b}{Sc} \zeta \delta_\Gamma           \\
    \label{eqn:dimless-momentum-lic-liq}
    \ddt{\phi U} + \div \phi U \otimes U                    &=
            - \grad \phi  P + \div \phi S + Gr \zeta \hat{g}
            + M_l                                           \\
    \label{eqn:dimless-momentum-lic-gas}
    \ddt{\phi \rho^* U} + \div \phi \rho^* U \otimes U      &=
            - \grad \phi P + \div \phi \mu^* S  + \rho^* Ar \hat{g}
            + M_g                                           \\
    \label{eqn:dimless-species-lic-liq}
    \ddt{\phi \zeta} + \div \zeta U   &=
              \frac{1}{Sc}\lap \zeta + \frac{Da_\Pi}{Sc} (\zeta + 1)
            + \frac{Sh_b}{Sc} \zeta                         \\
    \label{eqn:dimless-species-lic-liq}
    \phi_g \zeta                                            &= 1
\end{align}
subject to
\begin{align}
    \label{eqn:dimless-noslip}
    U \rvert_{wall} &= 0\\
    \label{eqn:dimless-flow}
    \grad \zeta \rvert_{wall} \cdot \hat{n}_{wall}
    &= Da (\zeta + 1)
\end{align}
where
$Gr$ is the Grashoff number,
$Ar$ is the Archimedes number,
$Sc$ is the Schmidt number,
$Da$ is the (first) Damkh\"{o}ler number,
and
$Da_\Pi$ is the second Damkh\"{o}ler number.

This body of equations rules the overall mass transfer phenomena occurring in a 
dissolved/dispersed system.
Let's explore the role of several terms in further detail:
The Grashoff number ($Gr$) represents the buoyancy effects of the liquid phase 
due to the density gradients induced by the concentration gradients.
However, this buoyancy effect is rapidly exceed by bubble-induced buoyancy.
This is produced by the density difference between the bubble and its 
surrounding liquid and fueled by gravity.
The intensity of this effect is represented by the Archimedes ($Ar$) number.

\section{Filtering}

The solution of the aforementioned equations is know to
exhibit very fine spatio-temporal features.
These are mainly due to two main physical process:
\begin{enumerate*}[label=(\roman*)]
    \item the chaotic (turbulent) nature of the solution,
        but also due to
    \item small-scale flow dynamics
\end{enumerate*}
However,
in most engineering applications we are interested in the large scale features.
Mathematically,
we can formulate these large scale features as a filtering of the solution.
We are then interested in obtaining filtered form of the governing equations,
which (ideally) is expressed in terms of the filtered variables only.

To do so, we introduce the filtering operator
\begin{equation}
    \filter{\cdot} = \frac{1}{\Omega}\int_\Omega \cdot d\Omega
    \label{eqn:filter}
\end{equation}
where $\Omega$ could be a characteristic time, space or thermodynamic state,
giving rise to time- space- or ensemble-averaged filters.
After filtering,
one can perform a Reynolds decomposition of any variable
as an average value plus an oscillation from that average
\begin{equation}
    u = \rav{u} + \ros{u}
    \label{eqn:Reynolds-decomposition}
\end{equation}
subject to $\filter{\ros{u}} = 0$.

By filtering the governing 
equations~\ref{eqn:dimless-mass-lic-gas}-\ref{eqn:dimless-species-lic-liq}, we 
obtain:

\begin{align}
    \label{eqn:dimless-mass-lic-gas}
    \ddt{\epsilon_l} + \div \epsilon_l \fav{U}                          &=
            - \frac{\rho_g}{\rho_l} \frac{Sh_b}{Sc}
              \epsilon_l \rav{\zeta}
              \delta_\Gamma
            - \boxed{\div \rav{\ros{\phi_l} \ros{u}}}                   \\
    \label{eqn:dimless-mass-lic-liq}
    \ddt{\epsilon_g} + \div \epsilon_g \fav{U}                          &=
            + \frac{Sh_b}{Sc}
              \epsilon_g \fav{\zeta}
              \delta_\Gamma
            - \boxed{\div \rav{\ros{\phi_g} \ros{u}}}                   \\
    \label{eqn:dimless-momentum-lic-liq}
    \ddt{\epsilon_l \fav{U}} + \div \epsilon_l \fav{U} \otimes \fav{U}  &=
            - \grad \epsilon_l  \rav{P}
            + \div \epsilon_l \fav{S}
            + Gr \rav{\zeta} \hat{g}
            + \rav{M_l}
            - \boxed{\div \epsilon \rav{\ros{u}\otimes\ros{u}}}         \\
    \label{eqn:dimless-momentum-lic-gas}
    \rho^* \ddt{\epsilon_g \fav{U}} + \div \epsilon_g \fav{U} \otimes            
            \fav{U}                                                     &=
            - \grad \epsilon_g \rav{P}
            + \mu^* \div \epsilon_g \fav{S}
            + \rho^* Ar \hat{g}
            + \rav{M_g}
            - \boxed{\div \epsilon \rav{\ros{u}\otimes\ros{u}}}         \\
    \label{eqn:dimless-species-lic-liq}
    \ddt{\epsilon_l \rav{\zeta}} + \div \epsilon_l \rav{\zeta} \fav{U}  &=
              \frac{1}{Sc}\lap  \epsilon_l \rav{\zeta}
            + \frac{Da_\Pi}{Sc} \epsilon_l (\rav{\zeta} + 1)
            + \frac{Sh_b}{Sc}   \epsilon_l \rav{\zeta}
            - \boxed{\div \epsilon \rav{\ros{\zeta}\ros{u}}}            \\
    \label{eqn:dimless-species-lic-liq}
    \epsilon_g \rav{\zeta}                                              &= 1
\end{align}
where we have introduced new terms (boxed) arising from the oscillatory 
behavior of the solution.
These terms require a sub-grid scale model in order to close the relations.

Let's analyze them in detail:

\begin{itemize}
    \item[$\rav{\ros{u} \otimes \ros{u}}$]
        this terms is the \textbf{Reynolds stress tensor} and describes the 
        additional stress experienced by the flow due to the turbulent 
        oscillations.
        The most common way to model these terms are the \textbf{eddy viscosity 
        models}, which assimilate them with an additional diffusion term.
    \item[$\rav{\ros{\zeta}\ros{u}}$]
        this terms is analogous to the turbulent dispersion force.
        However, it has two contributions: the purely turbulent dispersion,
        and the effect of the micromixing produced in the vicinity of the wall.
        Assuming the Reynolds analogy, we can assume that the turbulent 
        dispersion is proportional to the eddy viscosity,
        and so can be seen as the modification of the local Schmidt number.
        Regarding the micromixing effects, we need to come up with an additional 
        closure term to model them.
\end{itemize}

\subsection{Modeling of sub-scale terms}

\textbf{Mass transfer}

We will assess here the contribution of sub-scale terms to the mass transfer 
equation.
In particular, we will look at the filtered term of 
equation~\eqref{eqn:dimless-species-lic-liq} $\rav{\ros{\zeta} \otimes 
\ros{u}}$.
This term includes two different contributions which are very different in its 
nature.

The first contribution is \textbf{micromixing},
which is due to a perturbation within the mass transfer boundary layer.
When the solution is not locally isotropic within a computational domain
(e.g., near a wall)
filtering will result smear the solution,
introducing a non-physical damping of the solution.
In this cases,
we need to advance the resolution of the computation by so-called wall models.

The second contribution is \textbf{turbulent dispersion} due to the turbulent 
oscillations of the system.
On the other hand,
even when the solution is oscillatory within the computational domain,
since non--linear terms
(e.g., convection)
do not commute with filtering,
result into additional terms due to the propagation of the oscillations.
Such additional terms need to be modeled by what we call sub-grid scale models.
This term can be modeled analogously to the Reynolds stress term reported before 
by means of and eddy viscosity model.

\textbf{Wall model}
Since the typical size of mass transfer boundary layers is much smaller than the 
hydrodynamic ones, since they scale as:
\begin{equation}
    \frac{\delta}{\delta_c} = Sc^{-1/3}
    \label{eqn:mass-transfer-boundary-layer-scaling}
\end{equation}
for typical values of Schmidt numbers $Sc >> 1$ we obtain $\delta_c << \delta$.
Since the resolution required to resolve $\delta$,
the hydrodynamic boundary layer,
is already very high,
we can model the mass transfer boundary by means of a Sherwood-like expression 
as:
\begin{equation}
    \nabla \zeta \rvert_w = Sh_w c
    \label{eqn:mass-transfer-wall-model}
\end{equation}
which allows to work with a much coarser mesh.
\begin{figure}[h]
    \centering
        \begin{tikzpicture}
            \begin{axis}[xlabel = {$x$}, xmin = 0, ymin = 0, ymax = 1.5,
                         domain=0:1, clip mode = individual,
                         declare function = {d(\x) = 5*\x^2 - 3;}]
                \addplot[ultra thick] coordinates
                    {(0,0.25) (0.1, 1) (1, 1)};
                \addplot[] {d(x)};
                \draw[->]   ($(axis cs:0.00,1.1)$) --
                            node[above] {$\delta_c$}
                            ($(axis cs:0.10,1.1)$);
                \draw[dashed]   ($(axis cs:0.10,1.1)$) --
                                ($(axis cs:0.10,1.0)$);
                \draw[->]   ($(axis cs:0.00, {d(0.93)} )$) --
                            node[above] {$\delta$}
                            ($(axis cs:0.93, {d(0.93)} )$);
                \draw (axis cs:0.00,0.25) 
                node[fill,circle,scale=0.4,label={south east}:{$c_w$}]{};
                \draw (axis cs:0.50,1.00) 
                node[fill,circle,scale=0.4,label={south east}:{$c_i$}]{};
                \draw[<->]  (axis cs:0.00, -0.25) --
                            node[above] {$\Delta x$}
                            (axis cs:0.50, -0.25);
                \draw[dashed]   (axis cs:0.0,-0.10) --
                                (axis cs:0.0,-0.25);
                \draw[dashed]   (axis cs:0.50,0.25) --
                                (axis cs:0.50,-0.25);
                \draw   (axis cs:0.00,0.25) --
                        (axis cs:0.50,1.00);
                \draw[dashed] (axis cs:0.00,0.25) --
                              (axis cs:0.50,0.25);
                \draw[<->]   (axis cs:0.50,0.25) --
                node[fill=white, midway, align=center] {$c_i-c_w$}
                             (axis cs:0.50,1.00);

            \end{axis}
        \end{tikzpicture}
    \caption{The concentration profile around the eletrode wall exhibits a 
        really thin boundary layer ($\delta_c$),
        which is much thinner than the Blasius hydrodynamic boundary layer 
        ($\delta$),
        and the characteristic filter length ($\Delta x$).
        Obtaining the diffusive flow by a classical finite difference scheme 
        will result into an artificially low diffusion flux,
        since $\Delta c / \Delta x << \Delta c / \delta_c$,
        so we typically tune the value of $D$ by means of a wall model.}
    \label{fig:wall-model}
\end{figure}
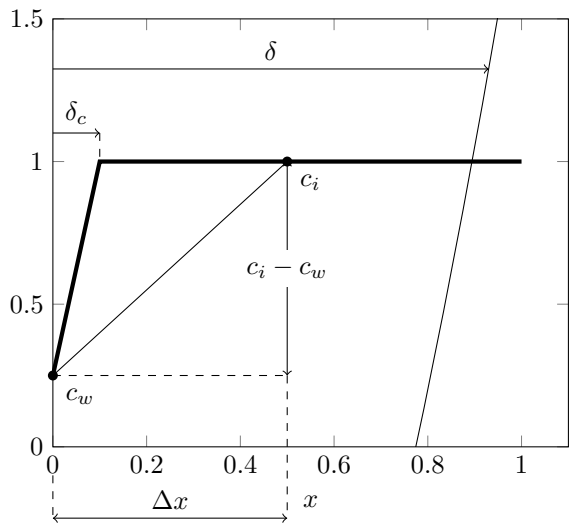

While a number of relations for $Sh$ exist for vertical planes,
the presence bubbles detaching from the electrode surface intensify the mass 
transfer of species towards the bulk.
Jansen et al. (1989) reported on the effect of gas bubbles,
distinguishing between two main mechanisms of bubble detachment:
\begin{enumerate*}[label=(\roman*)]
    \item bubble sliding, detachment and rising; and
    \item bubble sliding, coalescing and jumping.
\end{enumerate*}
The former corresponds with bubbles that typically do not coalesce,
like it has been reported for $H_2$,
which slide due to gravity over the electrode wall before departing into the 
bulk.
The later corresponds with bubbles that do nucleate,
like it has been reported for $O_2$,
which tend to coalesce violently on the electrode surface and violently jump 
into the bulk of the channel.

The data reported there can expressed as the superposition of the average $Sh$ 
number at a bubble-free wall,
plus an enhancement factor produced by the bubbles.
This model has obtained an outstanding match with experimental results,
as can be seen in Janssen1989.

We obtain,
for sliding and rising bubbles:
\begin{equation}
    Sh_{w,b} = Sh_w \left( 1 +
        v_{g \to b} \left(
            \frac{6\alpha_1^3s}{4\pi^2D}
                    \right)
    \right)
           + \frac{3\alpha_2^2R}{4\pi D Sh_w}
    \label{eqn:Sh-Janssen1989}
\end{equation}

\textbf{Sub-scale model}

\section{Boundary layer}
Boundary layers occur in the vicinity of the wall, where viscosity is dominat.
This allows to take several simplifications on the governing equations,
which enables some analytical incursion in the flow solutions.
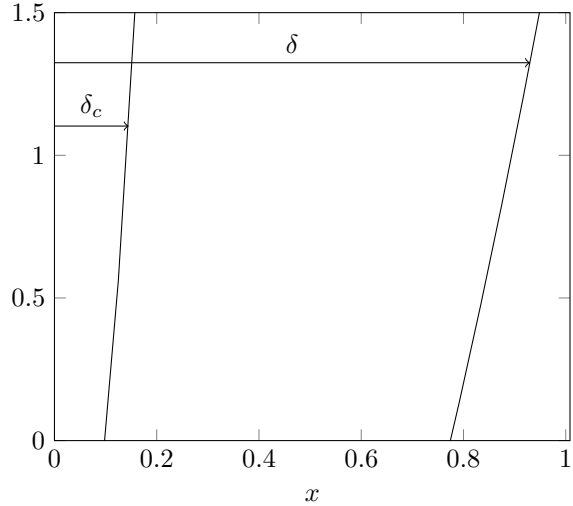
\begin{figure}[h]
    \centering
        \begin{tikzpicture}
            \begin{axis}[xlabel = {$x$}, xmin = 0, ymin = 0, ymax = 1.5,
                         domain=0:1, clip mode = individual,
                         declare function = {d(\x)  = 5*\x^2-3;
                                             dc(\x) = 100*\x^2-1;}]
                \addplot[] {d(x)};
                \addplot[] {dc(x)};
                \draw[->]   ($(axis cs:0.0,{dc(0.145)})$) --
                            node[above] {$\delta_c$}
                            ($(axis cs:0.145,{dc(0.145)})$);
                \draw[->]   ($(axis cs:0.00, {d(0.93)} )$) --
                            node[above] {$\delta$}
                            ($(axis cs:0.93, {d(0.93)} )$);

            \end{axis}
        \end{tikzpicture}
    \caption{The concentration profile around the eletrode wall exhibits a 
        really thin boundary layer ($\delta_c$),
        which is much thinner than the Blasius hydrodynamic boundary layer 
        ($\delta$).}
    \label{fig:boundary-layer}
\end{figure}

\subsection{Prandtl boundary layer equations}
Ludwig's Prandtl model for the boundary layer equations assumes a
2D,
incompressible,
stationary flow in the vicinity of a plate.
This results into the following set of equations:
\begin{align}
    \label{eqn:Prandtl-incompressible}
    \dd[x]{u} + \dd[y]{v} &= 0\\
    \label{eqn:Prandtl-mom-x}
    u \dd[x]{u} + v \dd[y]{u} &= -\frac{1}{\rho} \dd[x]{p} + \nu \dd[y^2]{^2u}\\
    \label{eqn:Prandtl-mom-y}
    \dd[y]{p} &= 0
\end{align}
assume that the variations along the stream-wise direction are much smaller than 
the variation in the wall-normal one.

Note that
equations~\eqref{eqn:Prandtl-incompressible}-\eqref{eqn:Prandtl-mom-y} involve 
three unknowns ($u$, $v$ and $p$) and three equations.

To simplify the equation a little bit further, we may assume that pressure 
variation in the streamwise direction is either zero (as for the case of 
boundary plates) or just hydrostatic (i.e., $p = \rho g x$).

However, since the flow is incompressible, we can express velocities in terms of 
a single streamfunction, $\Psi$.
\begin{align}
    u &= \dd[y]{\Psi}   &   v&= -\dd[x]{\Psi}
    \label{eqn:streamfunction}
\end{align}
This results in the following, single-variable PDE:
\begin{align}
    \label{eqn:Prandtl-stream}
    \dd[y]{\Psi}\dd[x\partial y]{^2 \Psi}   -
    \dd[x]{\Psi}\dd[y\partial y]{^2\Psi}    &=
    \nu \dd[y^2 y]{^3\Psi} +b
\end{align}
with the following boundary conditions:
\begin{align}
    \left.u\right|_{y=0}         &=\left.\dd[y]{\Psi}\right|_{y=0}         = 0 &
    \left.v\right|_{y=0}         &=-\left.\dd[x]{\Psi}\right|_{y=0}        = 0 &
    \left.u\right|_{x\to \infty} &=\left.\dd[x]{\Psi}\right|_{x\to \infty} = 0
    \label{eqn:Prandlt-boundary-conditions}
\end{align}

\subsection{Transport of scalars}
The mass transport equation of a scalar within the boundary layer is:
\begin{equation}
    u \dd[x]{\theta} + v \dd[y]{\theta} = D \dd[y^2]{^2\theta}
    \label{eqn:Prandtl-transport}
\end{equation}
which can be expressed in terms of the streamfunction as:
\begin{equation}
    \dd[y]{\Psi} \dd[x]{\theta} - \dd[x]{\Psi} \dd[y]{\theta}
    =
    D \dd[y^2]{^2\theta}
    \label{eqn:Prandtl-transport-stream}
\end{equation}
with the following boundary conditions
\begin{align}
    \left.\Delta c\right|_{y = 0}           &= c_{wall}-c_\infty    &
    \left.\Delta c\right|_{y \to \infty}    &= 0                    \\
    \left.D\Delta c\right|_{y = 0}          &= q                    &
    \left.\Delta c\right|_{y \to \infty}    &= 0
    \label{eqn:Prandtl-transport-boundary-conditions}
\end{align}
for the Dirichlet or the Neumann case, respectively.

\section{Self-similarity solutions}

Equation~\eqref{eqn:Prandtl-stream} involves two variables ($x$ and $y$),
which complicates its solution by analytical or numerical methods.

Self-similarity solutions are solutions to PDEs in terms of a self-similarity 
variable, $\eta$ which in general it is a function of both $x$ and $y$.
In this way, we can say that the solutions of the system at any point in space 
(i.e., in terms of $x$ and $y$) are ``similar'' to the solution in terms of 
$\eta$.

This tool allows to reduce the number of variables, provided that a 
self-similarity variable can be found or guessed.

Since most of the times we do not know the exact form of the self-similarity 
function \emph{a priori},
we attempt generic functions for the self-similarity variables and functions,
substitute them in the formulation of the original PDE
and finally the generic function to make sure that the resulting PDE is a 
function of the self-similarity variables solely.
We typically do so by assuming some sort of polynomial function on $x$ and $y$.
The procedure is the following:

First,
we do assume a shape for the self-similarity variable $\eta$ and the 
self-similarity solution $\Psi$.
These functions can, in principle, be any analytic function.
As fluid dynamicists,
this is where our understanding of the physics of the problem,
funneled trough scaling arguments,
may lead to fine educated guesses on the shape of the equations.

On the one hand,
we assume that the solution is self-similar in the streamwise direction.
This can be hinted from the shape of the equations,
since it includes a purely convective term
(which has a purely transport character)
in the streamwise direction,
while diffusion
(which has a spreading character)
in the wall-normal direction.
Since we are looking at boundary layer flows,
we render $\eta = y/\delta$ to represent the ratio of the wall-normal direction 
$y$ to the thickness of the boundary layer $\delta$, being $\delta$ some 
function of the streamwise coordinate $x$.
We further hint,
that $\delta$ is proportional to some power of $x$.
As such,
our candidate $\eta$ looks like:
\begin{equation}
    \eta = Ayx^{-s}
    \label{eqn:self-similar-eta}
\end{equation}

The velocity field is self-similiar in the streamwise direction.
However, since velocity is the derivative of the streamfunction, $\Psi$,
this implies that $\Psi$ is a combination of
a self-similar function $f(\eta)$
and some power of $x$.
This renders our candidate $f$ as:
\begin{equation}
    f(\eta) = B \Psi x^{-t}
    \label{eqn:self-similariy-f}
\end{equation}

The actual values of $s$ and $t$ is what gives \emph{the scalings} of the 
solution.

While some candidates may exist for $s$ and $t$ from experimental fittings,
we cannot say much on its value \emph{a priori}.
As such,
we will then proceed to obtain a self-similarity solution for 
equation~\eqref{eqn:Prandtl-stream}.

We start by describing the following intermediate variables:
\begin{align*}
    u              &= \dd[y]{\Psi}
                    = \dd[y]{\eta}\dd[\eta]{\Psi}
                   && = \frac{A}{B}x^{t-s}f'\\
    -v             &= \dd[x]{\Psi}
                    = \dd[x]{\eta}\dd[\eta]{\Psi}
                   && = \frac{1}{B}x^{t-1}\left( tf - s\eta f' \right)\\
    \dd[x]{u}      &= \dd[x\partial y]{^2\Psi}
                   &&= \frac{A}{B}x^{t-s-1}\bigl( (t-s) f' - s\eta f''\bigr)\\
    \dd[y]{u}      &= \dd[y^2]{^2\Psi}
                   && = \frac{A^2}{B}x^{t-2s}f''\\
    \dd[y^2]{^2 u} &= \dd[y^3]{^3\Psi}
                   && = \frac{A^3}{B}x^{t-3s}f'''
    \label{eqn:stream-derivatives}
\end{align*}
where
we have adopted the prime notation $(')$ to denote differentiation with respect 
to $\eta$ for the sake of compactness.
Since $f$ is a function of $\eta$ only,
and all differentiation is taken with respect to $\eta$ solely,
there is no risk of confusion.

This produces:
\begin{equation}
    f''' + \frac{1}{\nu A B} x^{t+s-1} \left(t f f' - (t-s)f'^2 \right)
    =
    b\frac{B}{\nu A^3} x^{3s-t}
    \label{eqn:self-similiarity-momentum}
\end{equation}
with the following boundary conditions:
\begin{align}
    \left.u\right|_{y=0}                        &= 0 =
    \left.\frac{A}{B}x^{t-s}f'\right|_{\eta=0}  &
    \left.f'\right|_0                           &= 0    \\
    \left.v\right|_{y=0}                        &= 0 =
    -\left.\frac{1}{B}x^{t-1}\left( tf - s\eta f' \right)\right|_{\eta=0} &
    \left.f\right|_0                            &= 0    \\
    \left.u\right|_{x\to \infty}                &= 0 =
    \left.\frac{A}{B}x^{t-s}f'\right|_{x\to \infty} &
    \left.f'\right|_\infty                      &= 0
    \label{eqn:bc-self-similar-flow}
\end{align}

Introducing the scalar self-similarity variable:
\begin{equation}
    \theta(\eta) = C \Delta c x^{-u}
    \label{eqn:self-similiary-theta}
\end{equation}
introducing the following intermediate variables
\begin{align*}
    \dd[x]{c} & = \frac{1}{C}x^{u-1}\bigl( u \theta' - s\eta \theta'\bigr)\\
    \dd[y]{c} & = \frac{A}{C}x^{u-s}\theta'\\
    \dd[y^2]{^2 c} & = \frac{A^2}{C}x^{u-2s}\theta''
    \label{eqn:theta-derivatives}
\end{align*}
and replacing them in the streamfunction formulation of the mass transport 
equation~\eqref{eqn:Prandtl-transport},
we obtain:
\begin{equation}
    \theta'' +
    \frac{1}{DAB}x^{t+s-1}\left( t f \theta' - u f' \theta \right) = 0
    \label{eqn:self-similiarity-transport}
\end{equation}
with the corresponding boundary conditions:
\begin{align}
    \label{eqn:bc-self-similar-transport-Dirichlet}
    \left.\theta\right|_0       &= 1   &
    & &
    \left.\theta'\right|_\infty &= 0   \\
    \label{eqn:bc-self-similar-transport-Neumann}
    & &
    \left.\theta'\right|_0      &= 1   &
    \left.\theta'\right|_\infty &= 0
\end{align}

\subsection{Scalings}
From the previous results we can obtain the corresponding boundary layer 
scalings $s$, $t$ and $u$ as those that render 
equations~\eqref{eqn:self-similiarity-momentum} 
and~\eqref{eqn:self-similiarity-transport}
and its boundary conditions (eq.~\eqref{eqn:bc-self-similar-flow} and
\eqref{eqn:bc-self-similar-transport-Dirichlet} 
or~\eqref{eqn:bc-self-similar-transport-Neumann})
exclusively in terms of $\eta$, $f$ and $\theta$.

A summary of the scaling factors can be find in the following table:
\begin{table}[h!]
    \centering
    \begin{tabular}{|c|c|c|c|c|c|}
        \hline
        Author  &   b                        &   s   &   t   & u        \\
        \hline
        Blasius\cite{Blasius1950} &   0                        & $1/2$ & $1/2$ & 
        N/A      \\
        Ostrach\cite{Ostrach1953} & $\propto C^{-1}\theta$     & $1/4$ & $3/4$ & 
        0        \\
        Sparrow\cite{Sparrow1959} & $\propto C^{-1}\theta x^u$ & $1/5$ & $4/5$ & 
        $1/5$    \\
        \hline
    \end{tabular}
    \caption{<+Caption text+>}
    \label{tab:<+label+>}
\end{table}

\subsection{Coefficients}
Now,
we still have the freedom to define the coefficients $A$, $B$ and $C$.
While this is a matter of preference, we report here the most usual practice.
In particular,
since we have three unknowns ($A$, $B$ and $C$),
we can impose three new equations,
namely for the two terms in which they appear in 
equation~\eqref{eqn:self-similiarity-prototype} such that the expression is 
simplified:
\begin{align}
    \frac{1}{\nu A B}   &= \frac{1}{s}\\
    \frac{bB}{\nu A^3}  &= \theta\\
    \frac{AD}{C}        &= q                &   C &=\frac{1}{c_{wall}}
    \label{eqn:self-similarity-coefficients}
\end{align}
This removes the fractional coefficients in 
equation~\eqref{eqn:self-similiarity-momentum} 
and~\eqref{eqn:self-similiarity-transport}
\begin{table}[h!]
    \centering
    \begin{tabular}{|c|c|c|c|c|c|}
        \hline
        Author  & A   &   B   &   C (Dirichlet) & C (Neumann)\\
        \hline
        Blasius
            & $\sqrt{\frac{U}{\nu}}$
            & $\frac{1}{\sqrt{\nu U}}$
            & $\frac{1}{C_{wall}}$
            & $\frac{D}{q}\sqrt{\frac{U}{\nu}}$\\
        Ostrach
            & $\left(\frac{g\beta \Delta\theta_w}{4\nu^2}\right)^{1/4}$
            & $\left(4^2\nu^2g\beta\Delta\theta_w\frac{q}{D}\right)^{-1/4}$
            & $\frac{1}{C_{wall}}$
            & N/A \\
        Sparrow
            & $\left(\frac{1}{5}\frac{g\beta}{\nu^2}\frac{q}{D}\right)^{ 1/5}$
            & $\left(5^4 \nu^3        g\beta        \frac{q}{D}\right)^{-1/5}$
            & N/A
            & $\left(\frac{1}{5}\frac{g\beta}{\nu^2}
                                \left(\frac{D}{q}\right)^4 \right)^{1/5}$ \\
        \hline
    \end{tabular}
    \caption{Summary of self-similarity scalings and coefficients.}
    \label{tab:self-similarity-coefficients}
\end{table}

%

\textbf{Blasius equation}
\begin{align}
    \label{eqn:Blasius-similarity-eta}
    \eta = \frac{y}{\delta} = Ayx^{-1/2}\\
    \label{eqn:Blasius-similarity-f}
    f    = B \Psi x^{-1/2}
\end{align}
\begin{equation}
    f''' + 2ff'' = 0
    \label{eqn:Blasius}
\end{equation}
\begin{align}
    \label{eqn:Blasius-boundary-conditions}
    f'(0) &= 0 & f(0) &=0 & f'(\infty) &=1
\end{align}

%
\textbf{Ostrach equation}
\begin{align}
    \label{eqn:Ostrach-similiarity-eta}
    \eta    &= \frac{y}{\delta} = Ayx^{-1/4}\\
    \label{eqn:Ostrach-similiarity-f}
    f       &= B \Psi x^{3/4}\\
    \label{eqn:Ostrach-similiarity-g}
    \theta  &= C \Delta c x^{-1/4}
\end{align}
\begin{align}
    <+content+>
    \label{<+label+>}
\end{align}<++>

%
%
%
%
%
\section{Leveque equation}
For the advection of passive scalars,
when $Sc >> 1$
(as it is typically the case in mass transfer)
the mass transfer boundary layer is much thinner than the hydrodynamic one.
This means that a linearization of the velocity profile is possible as:
\begin{equation}
    w(x) \approx \frac{\partial w}{\partial x} x = w' x
    \label{eqn:vel-linearization}
\end{equation}
This was observed by Leveque, who proposed a new scaling as:
\begin{equation}
        w' x \frac{\partial c}{\partial z} =
        D   \frac{\partial^2 c}{\partial x^2}
    \label{eqn:Leveque}
\end{equation}
where $w'$ and $D$ can be functions of $z$ only.

This last assumption is made in order to reuse Blasius-like expressions for 
$w'$, which then evolves in the streamwise direction $z$.

\subsection{Dirichlet boundary conditions}
To which we can attempt a self-similarity transformation as:
\begin{equation}
    \eta = Axf(z)^{-s}
    \label{eqn:Leveque-similarity-eta}
\end{equation}
where now $f(x)$ can be any function of x.
On the other hand, the actual variable $c$ is:
\begin{equation}
    c = c(\eta)
    \label{eqn:Leveque-similarity-c}
\end{equation}

By substituting these into the original equation~\eqref{eqn:Leveque}, we obtain:

\begin{equation}
    \frac{c''}{c'} =
    -s \frac{w'}{D} f' \frac{\eta^2}{A^{3}} f^{3s-1}
    \label{eqn:Leveque-raw}
\end{equation}
In order to satisfy self-similarity, we impose the following:
\begin{align}
    3s - 1          &= 0  \\
    \frac{w'}{D} f' &=1
    \label{eqn:Leveque-similarity-raw}
\end{align}
from where we obtain the final Leveque scaling:
\begin{align}
    s &= \frac{1}{3}\\
    f &= \int\frac{D}{w'} dz
    \label{eqn:Leveque-similarity}
\end{align}

With this scaling,
equation~\eqref{eqn:Leveque} becomes:
\begin{equation}
    \frac{c''}{c'} = -\frac{1}{A^{3}}\frac{1}{3}\eta^2\\
    \label{eqn:Leveque-fine}
\end{equation}
which is an eigenvalue problem.
We can reduce the order of the problem with realizing that
\begin{equation}
    \frac{c''}{c'} = \frac{\partial}{\partial \eta} ln(c')
    \label{eqn:log-derivative}
\end{equation}
Solving for it we obtain:
\begin{equation}
    c' = Cexp\left(-\frac{1}{A^{3}}\frac{\eta^3}{9}\right)
    \label{eqn:Leveque-sol}
\end{equation}
From where we can see that, for convenience, we can impose $A$ as:
\begin{equation}
    A = \left(\frac{1}{9}\right)^{1/3}
    \label{eqn:Leveque-similarity-constant}
\end{equation}
such that equation~\eqref{eqn:Leveque-sol}
\begin{equation}
    c' = C exp\left(-\eta^3\right)
    \label{eqn:Leveque-sol-gradient}
\end{equation}

Once $A$ is defined, the similarity is:
\begin{equation}
    \eta = y\left(\frac{1}{9\int\frac{D}{w'} dx}\right)^{\frac{1}{3}}
    \label{<+label+>}
\end{equation}

The final expression for $c$ is obtained by integration between $\eta$ and 
$\infty$.
\begin{equation}
    c_\infty - c(\eta) = C \int_\eta^\infty exp\left(-\eta^3\right) d\eta = C
    \label{<+label+>}
\end{equation}

\subsection{Neumann boundary conditions}

For Neumann boundary conditions:
\begin{align}
    D\frac{\partial c}{\partial y}\lvert_{y=0}  &= \frac{j}{zF} \\
    c\lvert_{y\to\infty}                        &= c_{sat}
    \label{<+label+>}
\end{align}

To which we attempt a self-similiarity solution of the form:
\begin{align}
    \eta            &= A y f(z)^{-s}\\
    \theta(\eta)    &= B \Delta c f(z)^{-u}
    \label{eqn:Leveque-Neumann-self-similarity}
\end{align}
where $\Delta c = c - c_sat$.
We scale $\theta$ as an arbitrary power of the shape function $f(x)$.
Substituting this scaling in the boundary conditions we obtain:
\begin{align}
    f^{u-s}\theta'(0)  &= \frac{j}{zFD}\frac{B}{A} \\
    \theta(\infty)             &= 0
    \label{<+label+>}
\end{align}
which requires $u-s = 0$ in order to obtain a self-similarity condition.
This means that $\eta$ and $\theta$ scale the same.

Introducing now the scaling in the Leveque equation~\eqref{eqn:Leveque},
we obtain the following ODE:
\begin{equation}
    w' \eta A^{-1} f^{s+u-1}f'\left( u \theta - s \eta \theta'\right)
    =
    D A^2 f^{-2s} B^{-1} f^u \theta''
    \label{<+label+>}
\end{equation}
which can be rearranged as
\begin{equation}
    \frac{w'}{D} f' f^{3s-1}\eta \left( u \theta - s \eta \theta' \right)
    =
    A^3 \theta''
    \label{<+label+>}
\end{equation}
from where we can impose the following self-similarity conditions:
\begin{align}
    \frac{w'}{D} f' &= 1\\
    3s-1            &= 0
    \label{<+label+>}
\end{align}

Self-similarity requires then scaling and the shape function as
\begin{align}
    f' &= \int\frac{D}{w'}\\
    s  &= \frac{1}{3}
    \label{<+label+>}
\end{align}
which fully define the self-similarity.
The equation then takes the form
\begin{equation}
    3A^3 \theta'' + \eta^2\theta' - \eta\theta = 0
    \label{<+label+>}
\end{equation}
where $A$ can be chosen to simplify the system as
\begin{equation}
    A = \left(\frac{1}{9}\right)^{1/3}
    \label{<+label+>}
\end{equation}
which renders the final form of the Leveque equation with Neumann boundary 
conditions as:
\begin{equation}
    \frac{1}{3}\theta'' + \eta^2 \theta' - \eta \theta = 0
    \label{<+label+>}
\end{equation}
which has an analytical solution if we attempt a solution like:
\begin{equation}
    \theta(\eta) = \eta v(\eta)
    \label{<+label+>}
\end{equation}
which accepts an analytical solution of the form
\begin{equation}
    \theta(\lambda) =  C E_{\frac{4}{3}}\left( \lambda^3 \right)
                    =  C \lambda \Gamma \left(-\frac{1}{3}, \lambda^3 \right)
    \label{eqn:theta-sol}
\end{equation}
where $E_i$ is the exponential integral,
while $\Gamma$ is the incomplete gamma function.
Taking its boundary conditions we obtain:
\begin{align}
    \theta'(0)      &= \Gamma \left( -\frac{1}{3} \right) C\\
    \theta(\infty)  &= 0
    \label{<+label+>}
\end{align}
Thus, the final form of $\theta$ becomes:
\begin{equation}
    \theta(\eta) =
    \eta \frac{\Gamma \left( -\frac{1}{3}, x^3 \right)}{\Gamma \left( 
        -\frac{1}{3}\right)} = \eta Q \left( -\frac{1}{3}, x^3 \right)
    \label{<+label+>}
\end{equation}
where $Q$ is the regularized upper incomplete gamma function.

The final expression for mass then becomes:
\begin{equation}
    \Delta c_{wall} = \frac{q}{D} \theta(\eta)
    \left( 9 \int_0^z \frac{D}{w'}dt \right)^{1/3}
    \label{<+label+>}
\end{equation}
which yields the expression for mass transfer coefficient as:
\begin{equation}
    k_f = \frac{1}{\theta(0)} \left( 9 \int_0^z \frac{D}{w'}dt \right)^{-1/3}
    \label{<+label+>}
\end{equation}
Where the final Sherwood number is:
\begin{equation}
    Sh_f = \frac{k_f z}{D} = \frac{1}{\theta(0)}
    z \left( 9 \int_0^z \frac{D}{w'}dt \right)^{-1/3}
    \label{<+label+>}
\end{equation}

\bibliographystyle{unsrt}
\bibliography{library.bib}

\end{document}